# Two Phonon Interactions and Charge Density Wave in Single Crystalline VSe$_2$ Probed by Raman Spectroscopy

Juhi Pandey[1] and Ajay Soni*

[1]School of Basic Sciences, Indian Institute of Technology Mandi, Mandi 175005, HP, India

*Email address: ajay@iitmandi.ac.in

Charge density wave (CDW) is a unique phenomenon mostly realized in two-dimensional (2D) metallic layered transition metal dichalcogenides (TMDCs). Here, we report on Raman spectroscopy of single crystal 1T-VSe$_2$ and observed signature of commensurate CDW (C-CDW) and incommensurate (I-CDW) transition, in the temperature range of 50-120 K. The room temperature Raman spectra showed a sharp $A_{1g}$ mode ~ 206 cm$^{-1}$ along with two new weak and broad Raman modes associated to $E_g$ mode ~ 257 cm$^{-1}$ and two-phonon mode ($2_{ph}$) ~ 332 cm$^{-1}$. The onset of I-CDW and C-CDW is estimated from resistance measurements supported by magnetic measurements. Remarkably, at the onset of I-CDW ~ 115 K, a significant enhancement in the intensity of weak $E_g$ mode is observed along with emergence of a doubly degenerate $E_g(2)$ mode ~ 144 cm$^{-1}$. Below 70 K, a weak $A_{1g}$ mode ~ 170 cm$^{-1}$ emerged signifying the onset of C-CDW. Anomalous phonon softening of $E_g$ mode ~ 257 cm$^{-1}$ and $2_{ph}$ process mode below I-CDW has been also observed, which is signifying the involvement of electron-phonon coupling.

Layered TMDCs provide a rich platform for research with their wide range of physical properties ranging from metallic, semiconducting, superconducting, magnetism along with exotic optical properties.[1] Most 2D materials exist in two stable crystal structures such as 2H (trigonal prismatic) and 1T (octahedron), where the physical properties depend on the symmetry of the phases.[2] Among the several versatile properties, CDW is a novel phenomenon realized in 1T and 2H polytypes of many metallic layered TMDCs, such as $NbSe_2$, $TaSe_2$, $TaS_2$, $VTe_2$, $VSe_2$, and has been received a rapid scientific and technological attention for many body interactions and potential applications in memory devices and oscillators.[3] CDW is a low-temperature ordered phase arising from periodic modulation of conduction electron densities accompanied by periodic distortion of atomic lattice.[4] The origin of CDW in TMDCs have not achieved any general scientific consensus and remained a topic of investigation so far.[5,6] The commonly accepted argument for origin of CDW is based on Kohn's anomaly where the electronic charge density is modulated periodically, which is represented with ordering vector, *Q*. The CDW develops an energy gap at the Fermi surface and thus Fermi surface nesting.[7] However, there are several recent reports suggesting that nesting alone is insufficient to explain the ordering vector, *Q*, especially for systems of 2D and higher dimensions.[6,8] In the case of $2H-NbSe_2$ and $ErTe_3$, the strong electron-phonon coupling plays a major role for determining the ordering vector than Fermi surface nesting.[9]

Based on the periodicity of *Q*, CDW has been classified as commensurate CDW (C-CDW), nearly commensurate (N-CDW) and incommensurate CDW (I-CDW). In C-CDW phase, the *Q* is integral multiple of reciprocal lattice vector of undistorted phase, whereas in I-CDW phase it is not. $1T-VSe_2$ shows I-CDW transition ~ 110 K and C-CDW transition ~ 80 K which have been studied using X-ray diffraction technique,[10] inelastic X-ray scattering technique, angle resolved photoemission spectroscopy,[11] transport measurement.[12] $1T-VSe_2$ is iso-structural with $1T-TaS_2$ and 1T-TaSe$_2$,[13] and expected to show similar anomaly in resistivity measurement at CDW, but unlike to these iso-structural compounds showing sharp anomaly in resistivity at CDW transitions,[14,15] VSe$_2$ shows a gradual upturn similar to 2H-NbSe$_2$, 2H-TaS$_2$ and 2H-TaSe$_2$.[16,17] Further, since electron-phonon coupling plays a major role in CDW compared to Fermi surface nesting, Raman spectroscopy has also been utilized to access CDW in 2H-NbSe$_2$, 2H-TaS$_2$ and 2H-TaSe$_2$.[18,19]

While the electronic structure has been well understood yet the phonon and their properties are still not well explored because of low signal and high reflectivity of 1T-VSe$_2$. Thus, we studied the CDW transitions in bulk single crystal of 1T-VSe$_2$, using temperature dependent resistivity, magnetic susceptibility and Raman spectroscopy measurements. The coupling between electron and phonon is demonstrated at low temperature in resistivity as well as Raman measurements, near CDW transitions.

**Experimental Section**

Single crystal of VSe$_2$ was grown by reacting stoichiometric amount of vanadium chips and selenium pellets (both from Sigma Aldrich, and 99.9%) using chemical vapor transport method with iodine as transport agent (~ 2 mg/cc). The reactants were vacuum sealed in quartz ampoule and placed in a two zone furnace. The two zones were heated slowly (~ 20 hrs) to reach growth temperature of 820 °C (hot zone) and 720 °C (cold zone). The samples were kept in the gradient temperature for ~ 72 hrs followed by slow cooling to room temperature in ~ 20 hrs. Thus, shiny large crystals of VSe$_2$ were obtained at the cold end of the tube. X-ray diffraction (XRD) of the prepared crystals was obtained using rotating anode Rigaku Smartlab diffractometer in Bragg-Brentano geometry with CuK$_\alpha$ radiation ($\lambda$ = 1.5406 Å). To determine the crystal structure and phase purity, Reitveld refinement of the XRD patterns was done using Fullprof Suite software.

Four probe electrical resistance ($R$) and magnetization ($\chi$) measurements were performed from 2 K to 300 K using Quantum Design make physical properties measurement system and magnetic properties measurement system, respectively. Micro-Raman measurements were done using Horiba Jobin-Yvon LabRAM HR evolution Raman spectrometer in back scattering geometry with solid state laser excitation (~ 532 nm), 1800 gr/mm grating and Peltier cooled CCD detector. Temperature dependent micro-Raman measurement was performed by using Montana cryostation in the temperature range of 3 K – 300 K. Field emission scanning electron microscopy (FESEM) images and elemental mapping using EDA were recorded through JFEI, USA make Nova Nano SEM-450 (supporting information).

**Result and Discussion**

The XRD pattern and Reitveld refinement of the VSe$_2$ is shown in FIG. 1 (a). The estimated lattice parameters are *a = 3.361* Å and *c = 6.101* Å, with a unit cell volume = 59.726 Å$^3$. The difference curve (Y$_{obs}$-Y$_{calc}$) clearly shows the grown VSe$_2$ crystals are of high purity single phase while the single crystalline nature is confirmed by observation of peaks only along the *c*-axis. Inset shows the optical image of the grown crystals with large dimensions (~ 5 mm x 5 mm). The bulk VSe$_2$ crystallizes in 1T polytype, where the V atoms are covalently bonded with the octahedra of Se atoms to form a single layer of VSe$_2$ and these layers are further stacked through van der Waals interactions. In general, the VSe$_2$ has trigonal crystal structure with $P\bar{3}m1$ space group (FIG. 1(b)), however, there is mild possibility of self-intercalation of V atoms between the van der Waals gap of VSe$_2$, thus occupying two different sites V1 (0, 0, 0) and V2 (0, 0, 0.5).[20] Such self-intercalation is commonly observed due to synthesis at high temperature.[21] Nevertheless, the extent of self-intercalation can be minimized by synthesizing the compound at lower temperature ~ 650 °C. The stoichiometry of the prepared sample has been estimated using elemental mapping (supporting information FIG. S1).

The temperature dependent electrical resistance ($R$) is shown in FIG. 2 (a), where metallic behavior has been observed throughout the $T$ range with an onset of the hump below 110 K. The hump is a signature of CDW transition, which is in-line with several layered TMDCs such as 1T-TaS$_2$, 1T-TaSe$_2$, 2H-TaSe$_2$.[14] The CDW transitions represent a close relationship between structural changes and the associated modifications of the charge density. The quality of crystal plays an important role in the onset of CDW transition and to assess the crystal quality, we determined the residual resistance ratio, $(RRR = R_{300K}/R_{6.2K})$. The $RRR$ of the sample is about 8 which is in good agreement with previous reports and confirms the high quality of synthesized VSe$_2$ single crystal.[12] The temperature range of the CDW transition is estimated from temperature derivative of resistivity, $dR/dT$, where two points of inflection at $T \sim 102$ K and $\sim 50$ K have been observed. In the study by Tsutsumi *et al.*,[22] the temperature dependent X-ray diffractions described the periodic lattice distortion wave vector in the reciprocal lattice as *(0.250 ± 0.003)a\* + (0.314 ± 0.003)c\** in I-CDW state below 110 K and *(0.250 ± 0.003)a\* + (0.307 ± 0.003)c\** in C-CDW state below 85 K. Here, the *a\** component is independent of temperature however, *c\** component shows deviation with temperature suggesting that the origin of incommensurability is along *c-axis*. Thus, the observed inflexions in $R$ measurement are thus associated to I-CDW ~ 102 K and C-CDW ~ 50 K.

The inset FIG. 2(a) shows low temperature region below 10 K, where $R$ increases with temperature exhibiting a minimum at ~ 6.2 K. Origin of such an upturn might result from weak localization, electron-electron interactions, or Kondo effect. The intercalated V$^{4+}$ ions with unpaired electron in *d* orbital provides a magnetic moments which can affect the electron transport at low temperatures.[21,23,24] Localized magnetic impurities can result in scattering of conduction electrons through s-d exchange interactions and the upturn observed in $R$ (or resisitivity ($\rho$)) through such interaction is described as Kondo effect.[25] The dependence of $\rho$ with $T$ including Kondo effect is represented with equation

$$\rho = \rho_0 + \alpha T^2 + c_m \ln\left(\frac{T_K}{T}\right) + bT^5 \qquad (1)$$

where $\rho_0$ is the residual resistivity, $\alpha T^2$ shows the contribution from the Fermi liquid properties, the term $bT^5$ is from the lattice vibrations and $c_m\left(\ln\frac{T_K}{T}\right)$ represents contribution from Kondo effect. The upturn of $R$ below 10 K is fitted with equation (1) suggesting the presence of Kondo effect. Recently, Barua et al.[25] have observed a similar upturn in $\rho$ of 1T-VSe$_2$ appearing due to V ions intercalated between the van der Waals gap which contributed to magnetic susceptibility and acted as localized magnetic scatters to the conduction electrons.

To confirm the presence of magnetic moments in 1T-VSe$_2$ because of intercalated V$^{+4}$ and its effect on CDW, the temperature dependent magnetic susceptibility ($\chi$) have been performed and the data are shown in FIG. 2(b). The $\chi$ shows a down turn at the onset of CDW in $R$ measurement. The inset in FIG. 2(b) shows the 1/$\chi$ from 1.8 K to 40 K (field ~ 100 Oe) fitted with a modified Curie Weiss law

$$\chi(T) = \frac{C}{T-\theta} + \chi_0 \qquad (2)$$

Where $\chi_0$ is a temperature independent intrinsic contribution to susceptibility from V bands, $C$ is a Curie Constant and $\theta$ is a paramagnetic Curie temperature. The parameters obtained from 1/$\chi$ fitting using modified Curie Weiss law are: $\chi_0 = 0.001313$ emu/mole-Oe, $C = 0.003013$ emu-K/mole-Oe, and $\theta = 0.07759$ K. Thus, the intercalated V is contributing to the feeble magnetism in the sample.

At high temperature above CDW transition, one formula unit of VSe$_2$ has three atoms per unit cell,[16,26] thus, has nine zero-center vibrational modes, which can be presented with irreducible representation, $\Gamma = A_{1g} + E_g(2) + 2A_{2u} + 2E_u(2)$. Here only two modes $A_{1g}$ and doubly degenerate $E_g(2)$ modes are Raman active while other four are IR active modes.[2] FIG. 3(a) shows Raman spectra at 300 K having three peaks at 206, 257 and 332 cm$^{-1}$ and all the modes are fitted with Lorentzian function. For VSe$_2$, Sugai *et. al.*[27] have reported that the sharp peak at ~ 206 cm$^{-1}$ is $A_{1g}$ symmetry. Due to the lack of group theoretical analysis of vibrational modes for 1T-VSe$_2$, the origin of the two new peaks ~ 257 cm$^{-1}$ and ~ 332 cm$^{-1}$ could not be assigned and thus requires further investigations. We have performed the polarized Raman spectra (FIG. S3, supporting information) to confirm the origin of these modes. The sharp $A_{1g}$ mode is observed only in perpendicular polarization while the mode ~ 257 cm$^{-1}$ is present in both parallel as well as perpendicular polarization measurements.[27] Thus, the mode at ~ 257 cm$^{-1}$ has $E_g$ symmetry. In low-dimensional conducting systems, such as highly anharmonic TMDCs, the anharmonicity plays an important role in formation of CDW through lattice distortion. For instance, in the Raman scattering an incident photon excites an electron-hole pair which can couple to an optic phonon. This phonon can decay to two-phonons with equal and opposite wave vectors through cubic anharmonicity of the system and resulting in a $2_{ph}$ peak.[28] The $2_{ph}$ process is a higher order process with weak intensities and it can carry the signature of Kohn anomaly phonon if the two wave vectors of $2_{ph}$ are $\pm 2k_F$. Broad and weak feature of $2_{ph}$ mode has also been observed above 300 cm$^{-1}$ for 1T-TiSe$_2$ through Raman spectroscopy.[29] Considering the fact that 1T-TiSe$_2$ has similar crystal structure and comparable molecular masses of V and Ti, the origin of the new modes can be consigned using the knowledge of vibrational modes of 1T-TiSe$_2$.[26,29] Therefore, the broad feature ~ 332 cm$^{-1}$ for 1T-VSe$_2$ can be related with two phonon process ($2_{ph}$) as assigned for 1T-TiSe$_2$.[16]

The low temperature (10 K) Raman spectra (FIG. 3(b)) has five Raman modes and is quite distinct from the 300 K. The peaks observed ~ 144 cm$^{-1}$, ~ 171 cm$^{-1}$, ~ 211 cm$^{-1}$, ~ 251 cm$^{-1}$ and 320 cm$^{-1}$ are assigned to doubly degenerate $E_g(2)$, $A_{1g}$, $A_{1g}$, $E_g$ and $2_{ph}$ modes, respectively.[30] Schematic of vibrational modes of $A_{1g}$ and $E_g(2)$ are shown in FIG. 3(c). Here, $A_{1g}$ mode represents out-of-plane vibration (along *c-axis*) of Se atoms, while the doubly degenerate $E_g(2)$ mode is associated with opposite vibration of Se atoms relative to each other along *a-axis* and *b-axis*.

To understand the role of CDW on phonons, we perform temperature dependent Raman study from 300 K to 10 K, as shown in FIG. 4. From 300 K to 120 K, the three Raman modes $A_{1g}$, $E_g$ and $2_{ph}$ remain exist and the intensity of $E_g$ and $2_{ph}$ is weak with broad spectral feature. The consistency between 120 K and 300 K Raman spectra confirms the stability of trigonal structure up to 120 K. However, below 120 K, the intensity of the broad $E_g$ mode starts to rise significantly and becomes comparable with the intensity of most intense $A_{1g}$ mode. Further, the doubly degenerate $E_g(2)$ mode starts to appear prominently between 100-10 K (shown in the red rectangle). Such modulation in the Raman spectra with temperature can be related with the onset of CDW transition below 120 K, as the size of the unit cell increases due to formation of superlattices.[27] Thus, the vibrational spectra is dramatically modulated by the formation of superlattices which have been probed by Raman and FTIR studies.[31] For instance, in the case of incommensurate distortions, the translational symmetry of the lattice is lost and the restriction of first order Raman scattering to zone center phonons is relaxed, resulting in participation of all of the modes of the material in first order Raman scattering.

Interestingly, a very minor peak of $A_{1g}$ ~ 171 cm$^{-1}$ appears at 40 K spectra and grows further in low temperature spectra (shown in green rectangle), as reflected in *R* measurement for the onset of commensurate phase below ~ 50 K. The behavior can be understood by the formation of commensurate superlattice where the size of Brillouin zone reduces, which is resulting in the folding of phonon dispersion curves. This can lead to appearance of new zone center phonons compared to undistorted structure. Therefore, the appearance of out-of-plane $A_{1g}$ mode below 50 K can be associated with formation of commensurate superlattice. Further, in earlier studies on X-ray investigation of superlattice at I-CDW and C-CDW transitions demonstrated the significant variation along $c^*$ axis at C-CDW transition whereas the component along $a^*$ axis remains independent of temperature, therefore the appearance of out-of plane $A_{1g}$ mode at ~ 50 K is a signature of C-CDW transition.

To explore the correlation between CDW transition with phonon energy and its intensity on temperature, the variation of peak position of $E_g(2)$ mode, $A_g$ mode ~ 211 cm$^{-1}$ and $E_g$ mode are shown in FIG. 5(a) and intensity of $E_g$ mode in FIG. 5 (b). The $E_g(2)$ mode and $A_g$ mode ~ 211 cm$^{-1}$ shows phonon hardening on lowering of temperature up to 10 K and has been fitted with $\omega(T) = \omega_0 + mT$, where $\omega_0$ is the phonon peak position/frequency of $E_g(2)$ and $A_g$ mode at absolute zero temperature, $m$ is the first order temperature coefficients of the two modes.[32] The fitted curve is shown as solid line in FIG. 5(a). The $m$ values for $E_g(2)$ and $A_g$ modes obtained by the linear fitting are -0.0617 and -0.0296 cm$^{-1}$/K respectively. The stiffening of phonon mode estimated from $m$ depends up on anharmonic coupling of phonons and thermal expansion of the crystal. In contrast to these two modes showing usual phonon hardening, the $E_g$ mode undergoes anomalous phonon softening with decrease in temperature as shown with solid blue line in FIG. 5(a). The strong softening of $2_{ph}$ shown in inset of FIG. 5(a) below I-CDW suggest the involvement of electron-phonon coupling at the onset of I-CDW for 1T-VSe$_2$. The decrease of phonon energy at $2k_F$ is related to generalized electronic susceptibility at $2k_F$ which results in large two phonon scattering intensity from Kohn anomaly modes.[33] Similar softening of the two phonon process is commonly observed in bulk and thin films of TaSe$_2$ in Raman spectra below I-CDW. [19] The intensity of $E_g$ mode which is active in both the CDW transitions enhances strongly below I-CDW transition temperature ~ 100 K as shown in FIG. 5(b) suggesting the coupling of Raman modes with CDW transitions.

**Conclusion**

In summary, the novel phenomenon of CDW has been investigated in single crystalline 1T-VSe$_2$ using temperature dependent resistance measurement and Raman spectroscopy. The hump in resistance measurement below 110 K is related to onset of I-CDW, while the existence of C-CDW ~ 50 K is confirmed by *dR/dT*. Appearance of five Raman active modes at low temperature suggests the reduction in size of Brillouin zone due to formation of superlattices. At the onset of I-CDW, the intensity of $E_g$ mode ~ 257 cm$^{-1}$ enhanced significantly and $E_g(2)$ mode appeared ~ 114 cm$^{-1}$. Softening of $E_g$ mode and $2_{ph}$ process below I-CDW signifies the role of electron-phonon coupling through Kohn anomaly in CDW transition.

**Acknowledgements:**

AS would like to acknowledge DST-SERB for funding (CRG/2018/002197 and YSS/2014/001038), and IIT Mandi, for research facilities. JP would like to acknowledge MHRD for providing fellowship.

**Figures and Figure Captions**

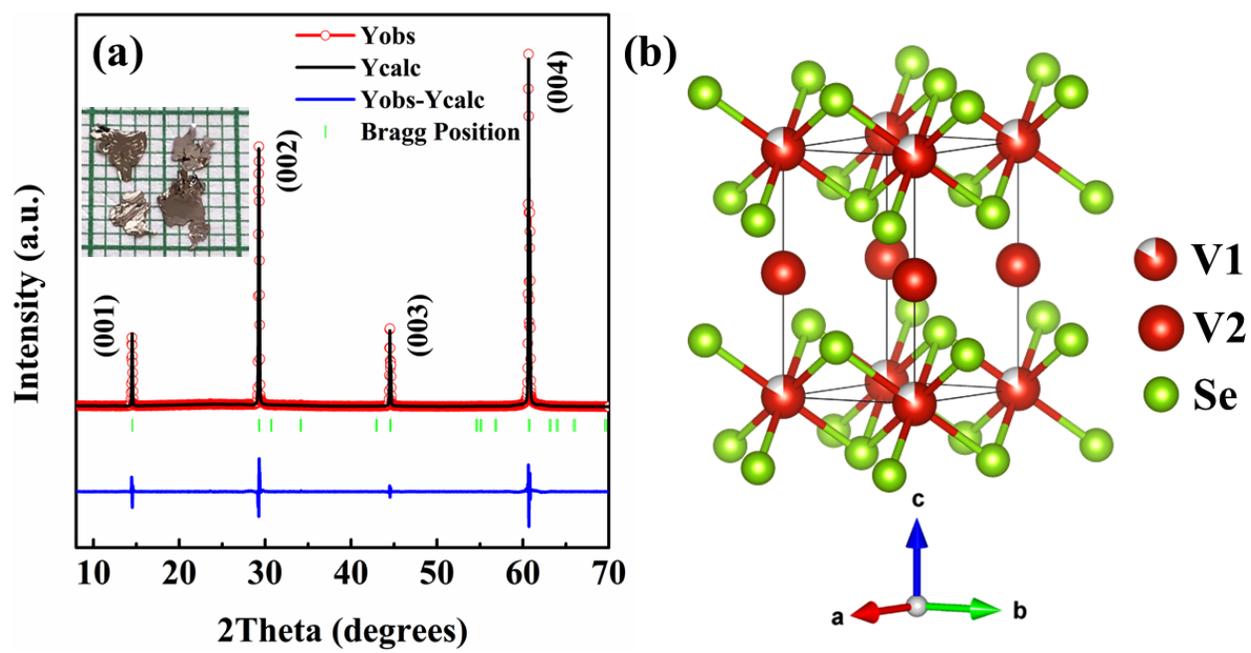

FIG. 1. (a) XRD pattern of single crystal of 1T-VSe$_2$, (b) schematic of trigonal crystal structure of $P\bar{3}m1$ space group.

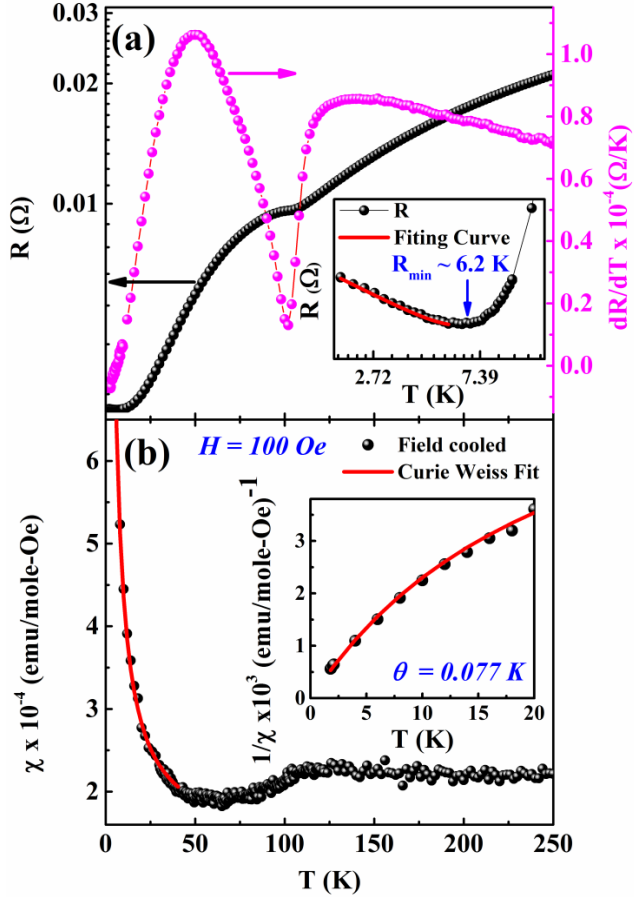

FIG. 2. (a) Resistance (*R*) and first derivative of resistance *(dR/dT)* of 1T-VSe$_2$; inset showing upturn in *R* below 6.2 K due to Kondo interactions. (b) Variation of magnetic susceptibility ($\chi$) with temperature showing downturn at the onset of CDW. The inset figure shows the variation of 1/$\chi$ with temperature fitted with modified Curie-Weiss law.

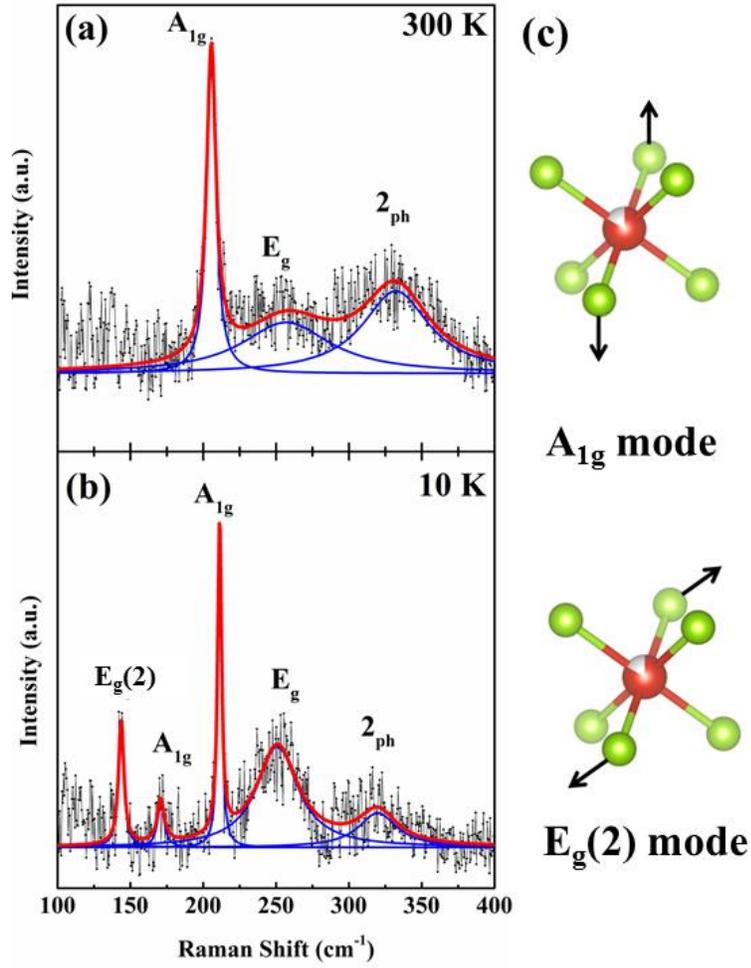

FIG. 3. Raman spectra of 1T-VSe$_2$ at (a) 300 K and (b) 10 K, and (c) Schematic of main type of lattice vibrations observed in Raman spectra.

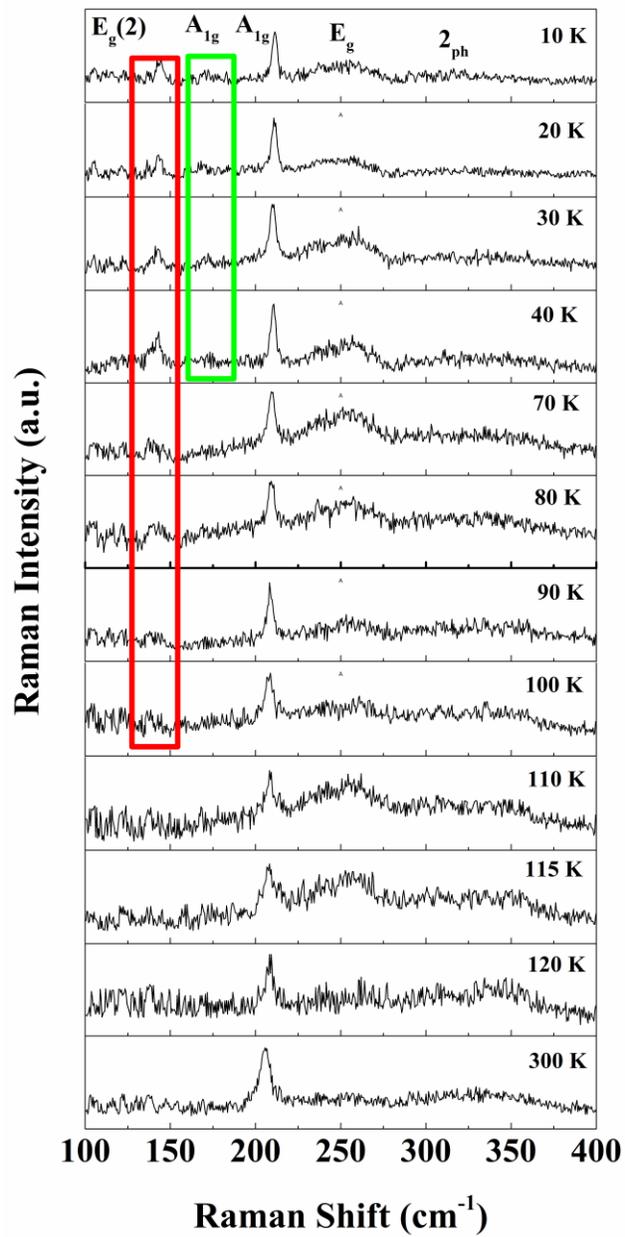

FIG. 4. Temperature dependence of Raman spectra from 300 K to 10 K.

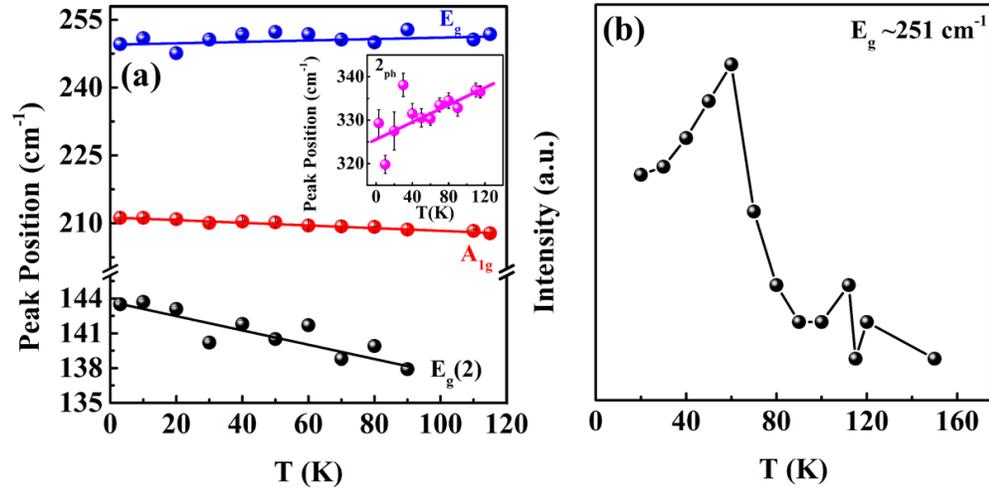

FIG. 5. Temperature dependent variation of (a) peak position of vibrational modes and (b) intensity of $E_g$ mode observed in the Raman spectra.